\documentclass{article}
\usepackage{spconf,amsmath,graphicx}
\usepackage{multirow}
\usepackage{hyperref}
\usepackage{subfigure}
\usepackage{graphicx}
\usepackage{url}
\usepackage{color}
\usepackage{verbatim}
\usepackage{balance}
\usepackage{amssymb}

\usepackage{booktabs}
\usepackage{caption}

\title{Multi-band MelGAN: Faster Waveform Generation for High-Quality Text-to-Speech}
\name{Geng Yang$^{1,2}$, Shan Yang$^1$, Kai Liu$^2$, Peng Fang$^2$, Wei Chen$^2$, Lei Xie$^{1*}$}
\address{
  $^1$Audio, Speech and Language Processing Group (ASLP@NPU), School of Computer Science, \\Northwestern Polytechnical University, Xi’an, China\\
  $^2$Sogou, Beijing, China}
  
\begin{document}
%
\maketitle
\begin{abstract}
In this paper, we propose multi-band MelGAN, a much faster waveform generation model targeting to high-quality text-to-speech. Specifically, we improve the original MelGAN by the following aspects. First, we increase the receptive field of the generator, which is proven to be beneficial to speech generation. Second, we substitute the feature matching loss with the multi-resolution STFT loss to better measure the difference between fake and real speech. Together with pre-training, this improvement leads to both better quality and better training stability. More importantly, we extend MelGAN with multi-band processing: the generator takes mel-spectrograms as input and produces sub-band signals which are subsequently summed back to full-band signals as discriminator input. The proposed multi-band MelGAN has achieved high MOS of 4.34 and 4.22 in waveform generation and TTS, respectively. With only 1.91M parameters, our model effectively reduces the total computational complexity of the original MelGAN from 5.85 to 0.95 GFLOPS. Our Pytorch implementation can achieve a real-time factor of 0.03 on CPU without hardware specific optimization.
\end{abstract}
\renewcommand{\thefootnote}{\fnsymbol{footnote}}
\footnotetext{* Lei Xie is the corresponding author. This work was supported by the National Key Research and Development Program of China (No.2017YFB1002102).}

\section{Introduction}
\label{sec:intro}

In recent years, neural network based waveform generation models have witnessed extraordinary success, which benefits tex-to-speech (TTS) systems with high-quality human-parity sounding, significantly surpassing speech generated with the conventional vocoders. Most high-fidelity \textit{neural vocoders} are autoregressive (AR), such as WaveNet~\cite{oord2016wavenet}, WaveRNN~\cite{kalchbrenner2018efficient},  SampleRNN~\cite{mehri2016samplernn}, etc. AR models are serial in nature, which relies on previous samples to generate current samples to model audio long-term dependencies. Although they can produce near-perfect wave samples, their generation efficiency is inherently low, which limits their practical use in efficiency-sensitive and real-time TTS applications.

AR models have been recently modified to speed up their inference~\cite{valin2019lpcnet,valin2019real,yu2019durian}. Two approaches are very competitive, both of which are variants of WaveRNN~\cite{kalchbrenner2018efficient}. In~\cite{yu2019durian}, a multi-band WaveRNN was proposed with over 2x speed-up in inference. A full-band audio was divided into four subbands, and by predicting the four subbands at the same time using the same network, the parameters of WaveRNN were significantly reduced. In~\cite{valin2019lpcnet, valin2019real}, the original WaveRNN structure was simplified by introducing linear prediction (LP), resulting in LPCNet. Combining LP with RNNs can significantly improve the efficiency of speech synthesis. 

Recently, significant efforts have been made to the development of non-AR models.  Because these models are highly parallelizable and can fully take advantages of modern deep learning hardware, they are extremely faster than their AR counterparts. One family relies on knowledge disillusion, including Parallel WaveNet~\cite{oord2017parallel} and Clarinet~\cite{ping2018clarinet}. Under this framework, the knowledge of an AR teacher model is transferred to a small student model based on the inverse auto-regressive flows (IAF)~\cite{kingma2016improved}. Although the IAF students can synthesize high-quality speech with a reasonable fast speed, this approach requires not only a well-trained teacher model but also some strategies to optimize the complex density distillation process.  The student is trained using a probability distillation objective, along with additional perceptual loss terms. In the meanwhile, such models rely on GPU inference in order to achieve a low real-time factor (RTF)\footnote{The real-time factor indicates the time required for the system to synthesize a one-second waveform, in seconds.} because of the huge amount of model parameters. The other family is flow-based models~\cite{kingma2018glow,dinh2014nice,dinh2016density}, including WaveGlow~\cite{prenger2019waveglow} and FloWaveNet~\cite{kim2018flowavenet}. They use a single network with the likelihood loss function only for training. As their inference process is parallel, the RTF is obviously lower as compared with the AR models. But it requires a week of training on eight GPUs to achieve good quality for a single speaker model~\cite{prenger2019waveglow}. While inference is fast on GPU, the large size of the model makes it impractical for applications with a constrained memory usage.

Generative adversarial networks (GANs)~\cite{goodfellow2014generative} are popular models for sample generation, which have been the dominating paradigm for image generation~\cite{gulrajani2017improved,karras2017progressive}, image-to-image translation~\cite{wang2018high} and video-to-video synthesis~\cite{wang2018video}. There were several early attempts applying GANs to audio generation tasks, but achieved limited success~\cite{yu2019durian}. Recently, there has been a new wave of modeling audio using GANs, as non-AR models targeting to fast audio generation. Specifically, MelGAN~\cite{kumar2019melgan}, Parallel WaveGAN~\cite{yamamoto2019parallel} and GAN-TTS~\cite{binkowski2019high} have shown promising performance on waveform generation tasks. They all rely on an adversarial game of two networks: a generator, which attempts to produce samples that mimic the reference distribution, and the discriminator, which tries to differentiate between real and generated samples. The input of MelGAN and Parallel WaveGAN is mel-spectrogram, while the input of GAN-TTS is linguistic features. Hence MelGAN and Parallel WaveGAN are considered as neural vocoders, while GAN-TTS is a stand-alone acoustic model. Meanwhile, Parallel WaveGAN and MelGAN both use auxiliary loss, i.e., multi-resolution STFT loss and feature matching loss, respectively, so they converge significantly faster than GAN-TTS. Impressively, the pytorch implementation of MelGAN runs at more than 100x faster than real-time on GPU and more than 2x faster than real-time on CPU. On the contrast, the real-time factor of Parallel WaveGAN is limited because of the stacking of network layers. According to the provided demos, the speech synthesized by MelGAN and Parallel WaveGAN is not satisfactory with audible artifacts.

In this paper, we propose a multi-band MelGAN (MB-MelGAN) for faster waveform generation and high-quality TTS. Specifically, we made several improvements on MelGAN to better facilitate speech generation. First, the receptive field has expanded to about twice of that in the original MelGAN, which is proven to be beneficial to speech generation, leading to obvious quality improvement. Second, we substitute the feature matching loss with more meaningful multi-resolution STFT loss as in Parallel WaveGAN, and combine with pre-training to further improve the speech quality and training stability. Third, to further improve the speech generation speed, we propose the multi-band MelGAN which can effectively reduce the computational cost. Similar to multi-band WaveRNN~\cite{yu2019durian}, we exploit the sparseness of neural network and adopt a single shared network for all sub-band signal predictions. Our study particularly shows that combing the sub-band loss with the full-band loss is beneficial to generation quality. The proposed MB-MelGAN, which has only 1.91M model parameters, effectively reduces the total computational complexity from 7.6 GFLOPS to 0.95 GFLOPS. Under the premise of obtaining 4.34 MOS, our Pytorch implementation can achieve a RTF of 0.03 on CPU without hardware specific optimization. The proposed MB-MelGAN vocoder further benefits end-to-end TTS with high quality speech generation performance. Audios can be found from: \url{http://yanggeng1995.github.io/demo}.

\section{The Model}
\label{sec:format}

\begin{figure}[t]
\centering
  \includegraphics[scale=1.0]{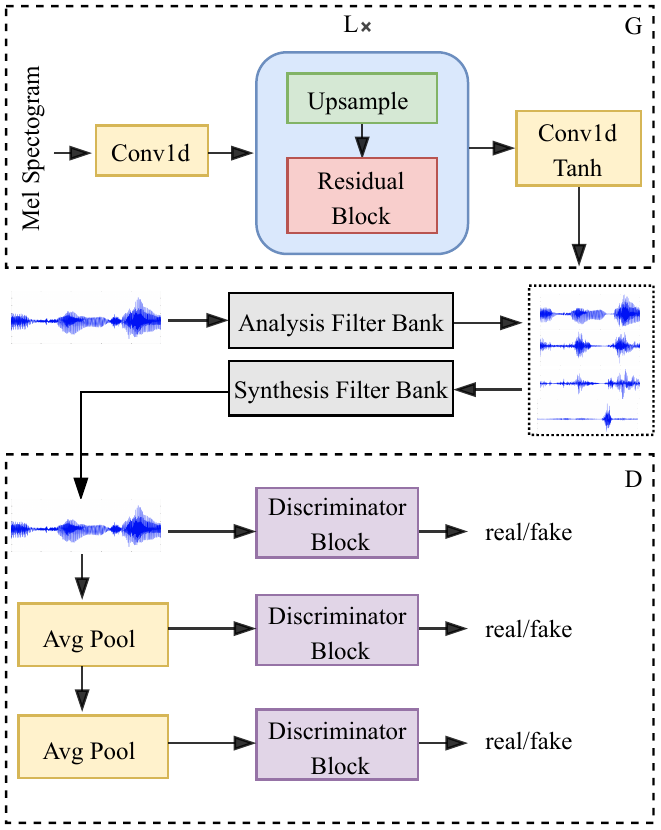}
  \caption{Multi-band MelGAN Architecture.}
  \label{fig:generator}
\end{figure}

Figure~\ref{fig:generator} illustrates the proposed multi-band MelGAN (MB-MelGAN), which is evolved from the basic MelGAN~\cite{kumar2019melgan}, following the general adversarial game between \textit{generator} and \textit{discriminator}. In MB-MelGAN, the generator network ($G$) takes mel-spectrogram as input to generate signals in multiple frequency bands instead of full frequency band in basic MelGAN. The predicted audio signals in each frequency band are upsampled first and then passed to the synthesis filters. The signals from each frequency band after synthesis filter are summed back to full-band audio signal. Then the discriminator network ($D$), in both basic MelGAN and the MB-MelGAN, treats full-band signal as input and use several discriminators to distinguish features originated from the generator in different scales.

\subsection{Basic MelGAN}
In the MelGAN generator~\cite{kumar2019melgan}, a stack of transposed convolution is adopted to upsample the mel sequence to match the frequency of waveforms. Each transposed convolution is followed by a stack of residual blocks with dilated convolutions to increase the receptive field, as shown in the upper box in Figure~\ref{fig:generator}. Using multiple discriminators is essential to the success of MelGAN, as single discriminator will produce metallic audio~\cite{kumar2019melgan}. Multiple discriminators at different scales are motivated from the fact that audio has fine-grained structures at different levels and each discriminator intends to learn features for different frequency range of audio. The multi-scale discriminators share the same network structure to operate on different audio scales in frequency domain. Specifically, with $K$ discriminators, basic MelGAN conducts adversarial training with objectives as:

  \begin{equation}
  \mathop{min}\limits_{D_k}\mathbb{E}_x\left[(D_k(x) - 1)^2\right] + \mathbb{E}_{s,z}\left[D_k(G(s, z))^2\right]
  \label{eq1}
  \end{equation}
\begin{equation}
\min _{G} \mathbb{E}_{s, z}\left[\sum_{k=1}^KD_{k}(G(s, z) - 1)^2\right]
\end{equation}
where $D_k$ is the $k^{th}$ discriminator, $x$ represents the raw waveform, $s$ means the input mel-spectrogram and $z$ indicates the Gaussian noise vector.

Basic MelGAN uses feature matching loss to minimize the $L1$ distance between the discriminator feature maps of real and synthetic audio at each intermediate layer of all discriminator blocks:
  \begin{equation}
  \mathcal{L}(G, D_k) = \mathbb{E}_{x,s}[\sum_{i=1}^T\frac{1}{N_i}\left \| D^{(i)}_k(x) - D^{(i)}_k(G(s))\right \|_1]
  \label{eq2}
  \end{equation}
where $D^{(i)}_k$ is the feature map output of the $i^{th}$ layer from the $k^{th}$ discriminator block, and $N_i$ is the number of units in each layer. Hence the final training objectives of basic MelGAN is:
  \begin{equation}
  \mathop{min}\limits_{G}(\mathbb{E}_{s,z}[\sum\limits_{k=1}^K(D_k(G(s, z)) - 1)^2] + \lambda\sum\limits_{k=1}^K\mathcal{L}(G, D_k)).
  \label{eq3}
  \end{equation}
\subsection{Proposed Multi-band MelGAN}


As discussed above, the original MelGAN uses latent features of the discriminator at different scales as a potential speech representation to calculate the difference between true and fake speech in Eq.~\eqref{eq2}. Although feature matching operation is helpful to stabilize the whole network, we find it is difficult to measure the differences between the potential features of true and fake speech, which causes the convergence process extremely slow. To solve this problem, we adopt the \textit{multi-resolution STFT loss}, which has been proven to be more effective to measure the difference between fake and real speech~\cite{yamamoto2019parallel,yamamoto2019probability,arik2018fast}.

For a single STFT loss, we minimize the spectral convergence $L_{sc}$ and log STFT magnitude $L_{mag}$ between the target waveform $x$ and the predicted audio $\widetilde{x}$ from the generator $G(s)$:
 \begin{equation}
   L_{sc}(x,  \widetilde{x}) = \frac{\left \|  \vert STFT(x)\vert - \vert STFT(\widetilde{x})\vert \right \|_F}{\left \|  \vert STFT(x)\vert\right \|_F}
  \label{eq4}
  \end{equation} 
  \begin{equation}
   L_{mag}(x,  \widetilde{x}) =\frac{1}{N}\left \| log\vert STFT(x)\vert - log\vert STFT(\widetilde{x})\vert \right \|_1
  \label{eq5}
  \end{equation}
  where $\left \|  \cdot\right \|_F$ and $\left \|  \cdot\right \|_1$ represent the Frobenius and $L_1$ norms, respectively. $|STFT(\cdot)|$ indicates the STFT function to compute magnitudes and $N$ is the number of elements in the magnitude. 
  
 For the multi-resolution STFT objective function, there are $M$ single STFT losses with different analysis parameters (i.e., FFT size, window size and hop size). We average the $M$ operations through
    \begin{equation}
  L_{mr\_stft}(G) = \mathbb{E}_{x, \widetilde{x}}\lbrack\frac{1}{M}\sum_{m=1}^M(L^m_{sc}(x, \widetilde{x}) +  L^m_{mag}(x, \widetilde{x}))\rbrack.
    \label{eq7}
  \end{equation} 
  
For the full-band version of our MelGAN (named as FB-MelGAN), we replace the feature matching loss with the multi-resolution STFT loss. Hence the final objective becomes
  \begin{equation}
  \mathop{min}\limits_{G}\mathbb{E}_{s,z}\lbrack\lambda\sum\limits_{k=1}^K(D_k(G(s, z)) - 1)^2\rbrack + \mathbb{E}_s\lbrack L_{mr\_stft}(G)\rbrack.
  \label{eq8}
  \end{equation}

For the multi-band MelGAN (MB-MelGAN), we conduct multi-resolution STFT in both full-band and sub-band scales. The final multi-resolution STFT of our MB-MelGAN becomes
  \begin{equation}
  L_{mr\_stft}(G) = \frac{1}{2}(L^{full}_{fmr\_stft}(G) + L^{sub}_{smr\_stft}(G))
  \label{eq9}
  \end{equation} 
  where $L^{full}_{{fmr}\_stft}$ and $L^{sub}_{{smr}\_stft}$ are the multi-resolution STFT loss in full-band and sub-band, respectively.
  
In detail, the proposed MB-MelGAN adopts a single generator for the prediction of all sub-bands signals. The shared generator takes mel-spectrogram as input and predicts all sub-bands simultaneously for sub-band multi-resolution STFT calculation, where the sub-band target waveforms are obtained through an analysis filter. Then we combine all sub-band audio signals into full-band scale through a synthesis filter to calculate full-band multi-resolution STFT loss with target full-band audio. We follow the method in~\cite{yu2019durian} to design the analysis and synthesis filters. Finally, we summarize the training procedure as follows.
\begin{itemize}
 \item[1)] Initialize $G$ and $D$ parameters;
 \item[2)] If \textbf{FB-MelGAN}, then pre-train $G$ using $L_{{mr}\_stft}(G)$ in Eq.~\eqref{eq7}, until $G$ converges;
 
 If \textbf{MB-MelGAN}, then pre-train $G$ using $L_{{mr}\_stft}(G)$ in Eq.~\eqref{eq9}, until $G$ converges;
 \item[3)] Train $D$ with Eq.~\eqref{eq1};
 \item[4)] Train $G$ with Eq.~\eqref{eq8};
 \item[5)] Loop 3) and 4) until the whole $G$-$D$ model converges.
 \end{itemize}
 Note that the $D$ network only presents in the model training, which is ignored in the waveform generation stage.

\section{Experiments}

\subsection{Experimental Setup}

For experiments, we use an open-source studio-quality corpus\footnote{Available at: \url{www.data-baker.com/open_source.html}} from a Chinese female speaker, which contains about 12 hours audio at 16kHz sampling rate. We leave out 20 sentences from the corpus for testing. We extract mel-spectrograms with 50 ms frame length, 12.5 ms frame shift, and 1024-point Fourier transform. The extracted spectrogram features are normalized to obey standard normal distribution before training. For evaluation, we adopt mean opinion score (MOS) tests to investigate the performance of the proposed methods. There are 20 native Chinese speakers evaluating the speech quality.

\subsection{Model details}
Table~\ref{tab:mb-melgan-architecture} shows the detailed structure of our improved MelGAN, for both full-band (FB) and multiband (MB) versions. They follow the general structure of basic MelGAN~\cite{kumar2019melgan} but with several modifications. We follow the method in~\cite{yu2019durian} for multi-band processing, in which a stable and efficient filter bank -- pseudo quadratue nirror filter bank (Pseudo-QMF) -- is adopted. Finite impulse response (FIR) analysis/synthesis filter order of 63 is chosen for uniformly spaced 4-band implementations. 

\textbf{Generator.} As for the upsampling module in our MB-MelGAN, 200x upsampling is conducted through 3 upsampling layers with 2x, 5x and 5x factors respectively because of predicting 4 sub-bands simultaneously, and the output channels of the 3 upsampling networks are 192, 96 and 48, respectively. Each upsampled layer is a transposed convolutional whose kernel-size being twice of the stride. FB-MelGAN has a slightly difference upsampling structure. Importantly, different from the basic MelGAN~\cite{kumar2019melgan}, we increase the receptive field by deepening the ResStack layers. We find that expanding receptive field to a reasonable size is helpful to improve the quality of speech generation, with a small model complexity increase but later compensated by introducing multi-band operations. Specifically, each residual dilated convolution stack (ResStack) has 4 layers with dilation 1, 3, 9 and 27 with kernel-size 3, having a total receptive field of 81 timesteps (in contrast with 27 in basic MelGAN~\cite{kumar2019melgan}). The output channel of the last convolution layer is 4 to predict 4-band audio or 1 to predict full-band audio.

\textbf{Discriminator.} FB-MelGAN and MB-MelGAN have the same discriminator structure which takes full-band audio (summed from 4 sub-band signals for MB-MelGAN) as input. Slightly different from the basic MelGAN, each discriminator block has 3 strided convolution (4 in basic MelGAN) with stride 4. We see no negative impact on performance with this simplification. Same with the basic MelGAN, we adopt a multi-scale architecture with 3 discriminators that have identical network structure but run on different audio scales: $D1$ operates on the scale of raw audio, while $D2$ and $D3$ operate on raw audio downsampled by a factor of 2 and 4 respectively. The multi-resolution STFT loss runs on different SFTF analysis parameters, as shown in Table~\ref{tab:stft-architecture}.

\begin{table}[t]
\footnotesize
\caption{Details of MB/FB-MelGAN model.}
\label{tab:mb-melgan-architecture}
\centering
\begin{tabular}{c|c|c|c}
\hline
Model &
  Layer &
  MB &
  FB \\ \hline
\multirow{5}{*}[-0.67cm]{Generator} &
  \begin{tabular}[c]{@{}c@{}}Conv1d (pad)\\ IReLU (0.2)\end{tabular} &
  7$\times$1, 384 &
  7$\times$1, 512 \\ \cline{2-4} 
 &
  \multirow{3}{*}[-0.35cm]{\begin{tabular}[c]{@{}c@{}}upsample\\ ResStack\\ IReLU (0.2)\end{tabular}} &
  \begin{tabular}[c]{@{}c@{}}$\times$2, 192\\ 192\end{tabular} &
  \begin{tabular}[c]{@{}c@{}}$\times$8, 256\\ 256\end{tabular} \\ \cline{3-4} 
 &
   &
  \begin{tabular}[c]{@{}c@{}}$\times$5, 96\\ 96\end{tabular} &
  \begin{tabular}[c]{@{}c@{}}$\times$5, 128\\ 128\end{tabular} \\ \cline{3-4} 
 &
   &
  \begin{tabular}[c]{@{}c@{}}$\times$5, 48\\ 48\end{tabular} &
  \begin{tabular}[c]{@{}c@{}}$\times$5, 64\\ 64\end{tabular} \\ \cline{2-4} 
 &
  \begin{tabular}[c]{@{}c@{}}Conv1d (pad)\\ Tanh\end{tabular} &
  7$\times$1, 4 &
  7$\times$1, 1 \\ \hline
\multirow{6}{*}[-0.1cm]{\begin{tabular}[c]{@{}c@{}}Discriminator\\ block\end{tabular}} &
  \multirow{5}{*}{\begin{tabular}[c]{@{}c@{}}Conv1d (pad)\\ IReLU (0.2)\end{tabular}} &
  \multicolumn{2}{c}{15$\times$1, 16} \\ \cline{3-4} 
 &
   &
  \multicolumn{2}{c}{41$\times$4, groups=4, 64} \\ \cline{3-4} 
 &
   &
  \multicolumn{2}{c}{41$\times$4, groups=16, 256} \\ \cline{3-4} 
 &
   &
  \multicolumn{2}{c}{41$\times$4, groups=64, 512} \\ \cline{3-4} 
 &
   &
  \multicolumn{2}{c}{5$\times$1, 512} \\ \cline{2-4} 
 &
  Conv1d (pad) &
  \multicolumn{2}{c}{3$\times$1, 1} \\ \hline
\end{tabular}
\end{table}

\begin{table}[t]
\caption{The parameters of multi-resolution STFT loss for full-band and multi-band, respectively. A Hanning window is applied before the FFT process.}
\label{tab:stft-architecture}
\footnotesize
\centering
\begin{tabular}{l|cccl}
\hline
                            & FFT size & Window size & Hop size &  \\ \hline
\multirow{3}{*}{Full-band}  & 1024     & 600         & 120      &  \\ 
                            & 2048     & 1200        & 240      &  \\ 
                            & 512      & 240         & 50       &  \\ \hline
\multirow{3}{*}{Multi-band}   & 384      & 150         & 30       &  \\
                            & 683      & 300         & 60       &  \\ 
                            & 171      & 60          & 10       &  \\ \hline
\end{tabular} 
\end{table}

\textbf{Training.} The initial learning rate of $G$ and $D$ is both set to $1e-4$ for all models for the Adam optimizer~\cite{kingma2014adam}. We also conduct weight normalization for all models. Model training is performed on a single NVIDIA TITAN Xp GPU, where the batch size for the basic/FB- MelGAN and MB-MelGAN is set to 48 and 128, respectively. Each batch randomly intercepts one second of audio. Since we find pre-training is effective for model convergence, we apply pre-training on the generator in the first 200K steps. The learning rate of all models is halved every 100K steps until $1e-6$. For models using feature matching loss, we set $\lambda$ = 10 in Eq.~\eqref{eq3}, while for models using multi-resolution STFT loss, we set $\lambda$ = 2.5 in Eq.~\eqref{eq8}.

\subsection{Evaluation}
\begin{table}[t]
\caption{The MOS results for different improvements on MelGAN (95\% confidence intervals). F in Index means Full-band.}
\label{tab:melgan-result}
\footnotesize
\centering
\begin{tabular}{c|l | c}
\hline
Index & Model             &MOS           \\ \hline
F0 & MelGAN~\cite{kumar2019melgan}   & 3.98$\pm$0.04  \\ 
F1 &  + Pretrain $G$ & 4.04$\pm$0.03 \\ 
F2 &  \ \ + $L_{mr\_stft}(G)$ & 4.06$\pm$0.04 \\ 
F3 &  \ \ \ \ + Deepen ResStack & \textbf{4.35$\pm$0.05}\\ \hline
\end{tabular}
\end{table}

\textbf{Improvements on basic MelGAN. } We first evaluated the proposed improvements on MelGAN which runs on full-band audio, as shown in Table \ref{tab:melgan-result}. System F0 shares the same architecture with the basic MelGAN in~\cite{kumar2019melgan}. With generator pre-training, we find system F1 outperforms the basic MelGAN (F0) with a small increase in MOS. Besides, we find the model converges much faster with pre-training -- training time is shortened to about two-third of basic MelGAN. When we further substitute the feature matching loss with the multi-resolution STFT loss, quality is further improved according to system F2. Another bonus is that the training time is further shortened to about one-third of basic MelGAN. Finally, by increasing the receptive field of system F2 to become system F3, we obtain a big improvement with the best MOS among all the systems. From the results, we can conclude that the proposed tricks about pre-training, multi-resolution STFT loss, and large receptive field are effective to achieve better quality and training stability. The listeners can tell audible artifacts such as jitter and metallic sounds in basic MelGAN (F0), while these artifacts seldomly appear in the improved versions, especially in system F3.

\textbf{Training strategy for MB-MelGAN. } Table \ref{tab:multi-band-melgan-result} shows the MOS results of the proposed MB-MelGAN. As previous evaluation shows the advantages of our architecture on the full-band version, in the multi-band version (MB-MelGAN), we follow the same architecture and tricks used in system F3. Firstly, we only use the multi-resolution STFT loss on the full-band waveform that is obtained from sub-band waveforms through the synthesis filter bank. We find this system, named M1, obtains a MOS of 4.22. We also notice that the introduction of multi-band processing lead to about 1/2 training time reduction as compared with the full-band models.  We further apply the  multi-resolution STFT loss directly on the predicted sub-band waveforms in system M2. The result shows that combining the sub- and full-band multi-resolution STFT losses is helpful to improve the quality of MB-MelGAN, leading to a big MOS gain. As an extra bonus, such combination can also improve training stability, leading to faster model convergence.

We notice that the final multi-band version (M2 in Table 4) has comparable high MOS with the improved full-band version (F3 in Table 3). We also trained multi-speaker version and 24KHz version of the proposed MB-MelGAN and evaluated the generation ability to unseen speakers as well. Informal testing shows the quality is pretty good. More samples can be found at: \url{https://yanggeng1995.github.io/demo}.

\begin{table}[t]
\caption{The MOS results for two training strategies on MB-MelGAN (95\% confidence intervals). M stands for multi-band.}
\label{tab:multi-band-melgan-result}
\footnotesize
\centering
\begin{tabular}{c|c|c|c}
\hline
Index & Model           &Loss  &MOS           \\ \hline
M1 & MB-MelGAN  & $L_{full}$ (Eq.~\eqref{eq7}) &4.22$\pm$0.04  \\ 
M2 & MB-MelGAN  & $L_{full} + L_{sub}$ (Eq.~\eqref{eq9}) & \textbf{4.34$\pm$0.03} \\ \hline
\end{tabular}
\end{table}

\textbf{Complexity. } We also evaluated the model size, generation complexity and efficiency, which are summarized in Table~\ref{tab:complexity}. Note that all the RTF values are measured on an Intel Xeon CPU E5-2630v3 using our PyTorch implementation without any hardware optimization. Our FB-MelGAN, indexed with F3, has a small noticeable increase in parameter size, computation complexity and real-time factor, mainly due to the enlargement of the receptive field. But its speech generation quality outperforms the basic MelGAN by a large margin (4.35 vs. 3.98 in MOS) according to Table 3. As for the proposed MB-MelGAN, we find it significantly decreases the model complexity, which generates speech about 7 times faster than basic MelGAN and FB-MelGAN. The most impressive conclusion is that MB-MelGAN retains the generation performance with a much smaller architecture and much better RTF.

\begin{table}[t]
  \caption{Model complexity.}
  \label{tab:complexity}
  \footnotesize
  \centering
\begin{tabular}{c|c|c|c|c}
\hline
Index & Model              & \multicolumn{1}{l|}{GFLOPS} & \multicolumn{1}{l|}{\#Paras. (M)} & \multicolumn{1}{l}{RTF}  \\    \hline
F0 & MelGAN~\cite{kumar2019melgan}   & 5.85   & 4.27 &0.2                         \\ 
F3 & FB-MelGAN       & 7.60   & 4.87 & 0.22                       \\ 
M2 & MB-MelGAN  & \textbf{0.95}   & \textbf{1.91} & \textbf{0.03}                        \\ \hline
\end{tabular}
\end{table}

\begin{table}[h]
\caption{The MOS results for TTS ($95\%$ confidence intervals).}
\label{tab:tts-result}
\footnotesize
\centering
\begin{tabular}{c|c|c|c}
\hline
TTS                        & Model     & Index & MOS  \\ \hline
\multirow{3}{*}{Tacotron2} & MelGAN~\cite{kumar2019melgan}   & F0    & 3.87$\pm$0.06 \\ 
                           & FB-MelGAN & F3    & 4.18$\pm$0.05 \\ 
                           & MB-MelGAN & M2    & \textbf{4.22$\pm$0.04} \\ \hline
\multicolumn{3}{c|}{Recording}                 & 4.58$\pm$0.03 \\ \hline
\end{tabular}
\end{table}

\textbf{Text-to-Speech.} In order to verify the effectiveness of the improved MelGAN as a vocoder for the text-to-speech (TTS) task, we finally combined the MelGAN vocoder with a Tacotron2-based~\cite{shen2018natural} acoustic model trained using the same training set. The Tacotron2 model takes syllable sequence with tone index and prosodic boundaries as input and outputs predicted mel-spectragrams which are subsequently fed into the MelGAN vocoder to produce waveform. Table \ref{tab:tts-result} summarizes the MOS values for the synthesized 20 testing sentences. The results indicate that the improved versions outperform the basic MelGAN by a large margin in the TTS task and the quality of the synthesized speech is the closest to the real recordings. Listeners can tell more artifacts (e.g., jitter and metallic effects) from the synthesized samples by the basic MelGAN, which is more severe in TTS as there exists inevitable mismatch between the predicted and the ground truth mel-spectragrams. On the contrast, the improved versions alleviate most of the artifacts with better listening quality.

\section{Conclusion}


This paper first presents some important improvements to the original MelGAN neural vocoder, which leads to significant quality improvement in speech generation, and further proposes a smaller and faster version of MelGAN using multi-band processing, which retains the same level of audio quality but runs 7 times faster. Text-to-speech experiments have also justified our proposed improvements. In the future, we will continue to fill the quality gap between the synthesized speech and the real human speech by improving GAN-liked neural vocoders.

\bibliographystyle{IEEEbib}
\bibliography{strings,refs}

\end{document}